\documentclass[11pt]{article}
\usepackage{amssymb}
\usepackage{amsmath}
\usepackage{graphics}
\usepackage{epsfig}
\usepackage{a4wide}
\usepackage{cite}
\usepackage{multirow}
\usepackage{color}
\usepackage{mathrsfs}
\usepackage{slashed}
\usepackage{subfigure}
\usepackage{booktabs}
\usepackage[utf8]{inputenc}

\DeclareGraphicsExtensions{.eps}
\begin{document}

\title{ Revisiting the determining fraction of glueball component in $f_0$ mesons via  radiative decays of $J/\psi$ }
\author{
Xing-Dao Guo$^{1,3}$\footnote{Corresponding author: guoxingdao@163.com}, Hong-Wei Ke$^2$, Ming-Gang Zhao$^3$ \\
Liang Tang$^4$, and Xue-Qian Li$^3$
\\ \\
{\small $^1$ College of Physics and New Energy, XuZhou University of Technology, Xuzhou 221111, Jaingsu, P.R. China}  \\
{\small $^2$ School of Science, Tianjin University, Tianjin, 300072, P.R. China,}  \\
{\small $^3$ Department of Physics, Nankai University, Tianjin 300071, P.R. China,}  \\
{\small $^4$ College of Physics, Hebei Normal University, Shijiazhuang 050024, P.R. China}
}

\date{}
\maketitle
\vskip 10mm

\begin{abstract}
QCD theory predicts the existence of glueballs, but so far all experimental endeavors have failed to identify any such states.
To remedy this discrepancy between QCD, which has proven to be a successful theory for strong interactions, and the failure of experimental searches for glueballs, one is tempted to accept the promising interpretation that the glueballs mix with regular $q\bar q$ states of the same quantum numbers.
The lattice estimate of the masses of pure $0^{++}$ glueballs  ranges from 1 to 2 GeV, which is the region of  the $f_0$ family.
Thus many authors suggest that the $f_0$ mesonic series is an ideal place to study possible mixtures of glueballs and $q\bar q$.
In this paper, following the strategy proposed by Close, Farrar and Li, we try to determine the fraction of glueball components in $f_0$ mesons using the measured mass spectra and the branching ratios of $J/\psi$ radiative decays into $f_0$ mesons.
Since the pioneering papers by Close et al., more than 20 years has elapsed and more accurate measurements have been done by several experimental collaborations, so it is time to revisit this interesting topic using new data.
We suppose $f_0(500)$ and $f_0(980)$ to be pure quark states, while for $f_0(1370)$, $f_0(1500)$ and $f_0(1710)$, to fit both the experimental data of $J/\psi$ radiative decay and their mass spectra, glueball components are needed.
Moreover, the mass of the pure $0^{++}$ glueball is phenomenologically determined.

\end{abstract}

\vfill \eject
\baselineskip=0.32in
\makeatletter      
\@addtoreset{equation}{section}
\makeatother       
\vskip 5mm
\renewcommand{\theequation}{\arabic{section}.\arabic{equation}}
\renewcommand{\thesection}{\Roman{section}.}
\newcommand{\nb}{\nonumber}

\section{Introduction}
\par
Quantum chromodynamics (QCD) theory demands the existence of glueballs because of interactions among gluons.
Glueballs behave differently from $q\bar q$ systems - for example, they do not directly couple to photons - so their special characteristics can help to identify glueball states.
Generally, several models based on lattice QCD\cite{Morningstar:1999rf, Chen:2005mg, Bali:1993fb} suggest $0^{++}$ glueball to be the lowest-lying glueball.
It has the same quantum numbers as the iso-singlet scalar meson $f_0$ family (i.e the so far observed series of $f_0(500)$, $f_0(980)$, $f_0(1370)$, $f_0(1500)$, $f_0(1710)$, $f_0(2020)$ and $f_0(2100)$).

\par
It is discouraging that after many years of exhaustive effort, no pure glueballs have been experimentally observed, even though theoretical studies repeatedly predict their existence and estimates of their masses have been presented.
Lattice QCD computations have predicted the mass of a $0^{++}$ glueball (which might be the lightest glueball) as $1.73$GeV\cite{Morningstar:1999rf}, $1.71$GeV\cite{Chen:2005mg} and $1.55\pm0.05$GeV\cite{Bali:1993fb}, while  the QCD sum rules determine it to be $1.50\pm0.2$GeV\cite{Narison:1996fm}, $1.71$GeV\cite{Huang:1998wj} and $1.50\pm0.06$GeV\cite{Shuiguo:2010ak}.
Phenomenological studies \cite{Close:1996yc, Amsler:1995td, Amsler:1995tu,Close:2000yk} suggest the mass of the lightest glueball to be around $1.5\sim 1.7$ GeV.
Moreover, in the Refs.\cite{Minkowski:2002nf, Harnett:2000fy, Forkel:2003mk} the authors suggest that the $0^{++}$ glueball might have two lower states \cite{Harnett:2000fy} with masses of $1.0\sim1.25$GeV and $1.4\pm0.2$GeV, and the authors of Ref.\cite{Forkel:2003mk} favor the mass of the $0^{++}$ glueball as $1.25\pm0.2$GeV.
Even though the theoretical estimates of the mass of the $0^{++}$ glueball are so diverse, they all suggest the mass to be within a range of 1.2 GeV$\sim$ 1.7 GeV.
A study of the mass of the $0^{++}$ glueball  based on analysis of the data would therefore be welcome.

\par
On other aspects, due to the failure of experimental searches for glueballs, we are tempted to consider that the QCD interaction would cause glueballs to mix with the $q\bar q$ states of the same quantum numbers, so that the possibility that pure glueballs exist independently in nature seems to be slim, even though it cannot be completely ruled out.
In other words, glueballs would mix with $q\bar q$ states to make hadrons.
This scenario could certainly reconcile the discrepancy between the QCD prediction and the experimental observation.
In fact, the mass of a pure glueball is only a parameter which does not have a definite physical meaning.

\par
In Refs.\cite{Amsler:1995td,Close:2005vf,Giacosa:2005zt,Janowski:2011gt,Cheng:2015iaa,Frere:2015xxa,Li:2000cj}, the authors considered $f_0(1370)$, $f_0(1500)$ and $f_0(1710)$ as mixtures of glueballs and $J^{PC}=0^{++}$ $q \bar q$ bound states.
They preferred $f_0(1500)$ as a hadron dominated by a glueball component.
Furthermore, in Ref.\cite{He:2006tw,He:2006ij} the authors further extended the scenario by involving possible components of the hybrid state $q\bar q g$ which may provide better fits to the available data.
Contrary to the above consideration, in Ref.\cite{Cheng:2015iaa,Frere:2015xxa}, $f_0(1710)$ was supposed to be dominated by the glueball component, but not a pure glueball.

\par
As the first step, in this work, we restrict ourselves to the scenario where only  mixtures of glueballs and $q\bar q$ are considered, while a possible contribution of hybrids to the mass spectra of the $f_0$ family is ignored.

\par
We first calculate the masses of the $q\bar q$ bound states by solving the Schr$\ddot{o}$dinger equations.
Some authors have extended the equation into its relativistic form for estimating the mass spectra of heavy-light mesons and the results seem to be closer to the data.
Following Ref.\cite{Liu:2013maa}, we calculate the light quark-antiquark system in the relativistic Schr$\ddot{o}$dinger equations.

\par
The results indicate that the experimentally measured masses of $f_0(500)$ and $f_0(980)$ can correspond to the $q\bar q$ states (ground states of $\frac{u\bar u + d\bar d}{\sqrt 2}$ and $s\bar s$), so can be considered as pure bound quark-antiquark states.
However, their spectra (including ground and excited states) do not correspond to the physical masses $f_0(1370)$, $f_0(1500)$ and $f_0(1710)$, signifiing that these cannot be pure $q\bar q$ states and extra components should be involved.
To evaluate the fractions of glueballs in those states, diagonalizing the mass matrix whose eigenvalues correspond to the masses of the physical states and the transformation unitary matrix determines the fractions of $q\bar q$ and glueball in the mixtures.
We define four parameters: $\lambda_{N-G},\; \lambda_{S-G},\; \lambda_{N-S}$ and $m_G$ which respectively are the mixing parameters between
$\frac{u\bar u + d\bar d}{\sqrt 2}$ and glueball, $s\bar s$ and glueball, $\frac{u\bar u + d\bar d}{\sqrt 2}$ and $s\bar s$ states,
and the mass of a pure glueball.
Even though fixing these four parameters can be done with this manipulation, to be more convincing and accurate, we adopt the the strategy provided by Close, Farrar and Li\cite{Close:1996yc}, analyzing the radiative decays of $J/\psi\to\gamma + f_0$ to reproduce those parameters, so that the results can be checked.

\par
After this introduction, in Section II we calculate the mass spectra of $q\bar q$ states of $0^{++}$ by solving the relativistic Schr$\ddot{o}$dinger equations.
In Section III, via a full analysis we confirm three physical states ($f_0(1370)$, $f_0(1500)$ and $f_0(1710)$) as  mixtures of $q\bar q$ and glueballs.
Then in the following section, we present the scheme of Close, Farrar et al. for $J/\psi\to \gamma+ f_0$ where $f_0$ refers to $f_0(1370)$, $f_0(1500)$ and $f_0(1710)$, and extract useful information about the fraction of glueball components in those mesons.
Then we calculate several ratios which may help to clarify the structures of various $f_0$ states.
A brief discussion and conclusion are presented in the last section.

\section{ $0^{++}$ $q\bar q$  systems }
\par
First, in terms of the relativistic Schr$\ddot{o}$dinger equation, let us calculate the mass spectra of $f_0$ by assuming them to be made of a light quark and an anti-quark.
Using the so-called relativistic Schr\"odinger equation is only an improvement to the regular one.
In the Hamiltonian, only the concerned kinetic part of the light quark(anti-quark) adopts the relativistic form and the other part is unchanged.
Because the light quarks are not as heavy as $c$ or $b$ quarks, one can believe that the modification may provide a physical picture which is closer to the physical reality.
But, of course, it is not like the Dirac equation; it is only an improvement of the non-relativistic Schr\"odinger equation.

\par
In Ref.\cite{Godfrey:2004ya} the authors study the $B_c$ meson, which contains two heavy quarks(anti-quarks) through the relativistic Schr\"odinger equation, while in Refs.\cite{Liu:2013maa,Colangelo:1998eb,Cea:1988yd} mesons which contain a heavy quark(anti-quark) and a light quark(anti-quark) were investigated in the same scenario.
In Ref.\cite{Cea:1982rg} the authors studied $\phi$ in terms of the relativistic Schr\"odinger equation and in Ref.\cite{Cea:1983wt} the authors studied $\rho$ and $\phi$ via the relativistic Schr\"odinger equation.
In Ref.\cite{Godfrey:1985xj} many light mesons with different quantum numbers have been studied in terms of the relativistic Schr\"odinger equation.
In all these studies, obvious improvements were reported, namely the resulting solutions are closer to the data.
Since $f_0$ mesons are $0^{++}$ states, the relative orbital angular momentum $l=1$.
Following Ref.\cite{Liu:2013maa}, the effective Hamiltonian is
\begin{equation}
\begin{array}{rl}
H =\sqrt{-\nabla^2_1+m_1^2}+\sqrt{-\nabla^2_2+m_2^2}+V_0(r)+H',
\end{array}
\end{equation}
where $m_1$ and $m_2$ are the masses of the light quark and anti-quark respectively.
In our numerical computations we set $m_1=m_2=0.3$ GeV for the $u$ and $d$ quark, and $m_1=m_2=0.5$ GeV for the $s$ quark. $\nabla^2_1$ and $\nabla^2_2$ act on the fields of $q_1$ and $\bar q_2$, $V_0(r)$ is a combination of the QCD-Coulomb term and a linear confining term\cite{Eichten:1978tg,Eichten:1979ms,Godfrey:1985xj}
\begin{equation}
\begin{array}{rl}
V_0(r) =\frac{-4}{3}\frac{\alpha_s(r)}{r}+\kappa r+c.
\end{array}
\end{equation}
Here $\alpha_s(r)$ is the coupling constant.
For the concerned energy scale of $\Lambda_{QCD}\sim 300$ MeV  the non-perturbative QCD effect dominates and so far $\alpha_s(r)$ cannot be determined by a general principle.
Thus, one generally, needs to invoke concrete models where the model-dependent parameters are adopted by fitting data.
Indeed, theoretical uncertainties are unavoidable.
In this work, the running coupling constant $\alpha_s(r)$, expressed in terms of a function of coordinates, can be obtained through the Fourier transformation of $\alpha_s(Q^2)$.
Following Refs.\cite{Liu:2013maa,Godfrey:1985xj}, we have
\begin{equation}
\begin{array}{rl}
\alpha_s(r)=\Sigma_i \alpha_i \frac{2}{\sqrt \pi}\int^{\gamma_i r}_0 e^{-x^2} dx
\end{array}
\end{equation}
where $\alpha_i$ and $\gamma_i$ are free constants, which were fitted \cite{Liu:2013maa,Godfrey:1985xj} by making the behavior of the running coupling constant $\alpha_s(r)$ at short distances coincide numerically with $\alpha_s(Q^2)$ predicted by QCD.
In our calculation we follow their work and take $\alpha_1=0.15,\; \alpha_2=0.15,\; \alpha_3=0.20$ and $\gamma_1=1/2,\; \gamma_2=\sqrt {10}/2,\; \gamma_3=\sqrt {1000}/2$.

\par
Since we are dealing with the P-wave structure of $q\bar q$, the spin-spin hyperfine interaction and spin-orbit interaction are concerned and an extra Hamiltonian $H'$ can be written as
\begin{equation}
\begin{array}{rl}
H' =V_{hyp}(r)+V_{so}(r).
\end{array}
\end{equation}
The spin-spin hyperfine interaction is
\begin{large}
\begin{equation}
\begin{array}{rl}
V_{hyp}(r) =\frac{32\pi}{9 m_1 m_2}\alpha_s\delta_\sigma(r)\mathbf{s}_1\cdot \mathbf{s}_2
-\frac{4}{3}\frac{\alpha_s}{m_1 m_2}
\frac{1}{ r^3}(\frac{3\; \mathbf{s}_1\cdot\mathbf{r}\; \mathbf{s}_2\cdot\mathbf{r} }{r^2}-\mathbf{s}_1\cdot \mathbf{s}_2),
\end{array}
\end{equation}
\end{large}
with\cite{Liu:2013maa}
\begin{equation}
\begin{array}{rl}
\delta_\sigma(r) =(\frac{\sigma}{\sqrt \pi})^3 e^{-\sigma^2 r^2},
\end{array}
\end{equation}
where $\sigma$ is a phenomenological parameter and $\langle\mathbf{s}_1\cdot\mathbf{s}_2\rangle=1/4$.

\par
The spin-orbit interaction is
\begin{large}
\begin{equation}
\begin{array}{rl}
V_{so}(r) =\frac{4}{3}\frac{\alpha_s}{ r^3}(\frac{1}{m_1}+\frac{1}{m_2})(\frac{\mathbf{s}_1\cdot\mathbf{L}}{m_1}+
\frac{\mathbf{s}_2\cdot\mathbf{L} }{m_2})-
\frac{1}{2r}\frac{\partial V_0(r)}{\partial r}(\frac{\mathbf{s}_1\cdot\mathbf{L}}{m_1^2}+\frac{\mathbf{s}_2\cdot\mathbf{L}}{m_2^2}),
\end{array}
\end{equation}
\end{large}
where $\mathbf{L}$ is the orbital angular momentum between the quark and anti-quark. For the $0^{++}$ state, we have $\langle\mathbf{s}_1\cdot\mathbf{L}\rangle= \langle\mathbf{s}_2\cdot\mathbf{L}\rangle =-1$.

\par
Determining the five parameters is a bit tricky.
We are dealing with $f_0$ mesons whose contents do not include heavy quarks, so when using the potential model to calculate their mass spectra we need to adopt different schemes from that for heavy quarkonia, to determine the relevant parameters.
Our strategy is as follows.
We suppose $f_0(500)$ and $f_0(980)$ are pure quark states, i.e. mixtures of $\frac{u\bar u + d\bar d}{\sqrt 2}$ (which we abbreviate as $n\bar n$) and $s\bar s$,  and try to evaluate the mass eigenvalues of $n\bar n$ and $s\bar s$ (which are not physical states).

\par
We then try to fix the other two parameters $\kappa$ and $c$.
We define $m_{0n\bar n}$, $m_{1n\bar n}$ and $m_{2n\bar n}$ as the masses of the ground state, first excited state and the second excited state of $n\bar n$.
Similarly, for $s\bar s$ we have $m_{0s\bar s}, m_{1s\bar s},m_{2s\bar s}$.
Since there are no precise data available, according to the analyses made by previous authors we can set several inequalities as:

$$m_{f0(500)}\leq m_{0n\bar n} < m_{0s\bar s}\leq m_{f_0(980)},$$
$$m_{f_0(1370)}\leq  m_{1n\bar n} < m_{1s\bar s}\leq  m_{f_0(1710)},$$
$$m_{f_0(2020)} \leq m_{2n\bar n} < m_{2s\bar s}\leq m_{f_0(2100)}.$$

By these criteria, we cannot obtain exact numbers for $b$ and $c$, but can set ranges for them.
Fortunately the ranges are not too wide for further phenomenological applications.
To satisfy the above constraints, we obtain $\kappa = 0.29 \sim 0.33$  GeV$^2$ and $c = -1.72 \sim -1.58$ GeV.
The masses of the ground, first excited and second excited states of $n\bar n$ and $s\bar s$ as are obtained as listed in Tab.\ref{tab1}.
From the table, we note that the masses of the ground states of $n\bar n$ and $s\bar s$ are respectively $626\sim 636 $ MeV and $830 \sim 848$ MeV.
We notice that all the achieved values are within certain ranges, but are not fixed numbers, as the discussed above.
\begin{table}[htbp]
\begin{minipage}[!t]{\columnwidth}
  \renewcommand{\arraystretch}{1.3}
  \centering
  \setlength{\tabcolsep}{0.6mm}
\begin{tabular*}{150mm}{c@{\extracolsep{\fill}}ccccc}\toprule[1pt]
principal quantum number    &     n=1                   &            n=2          & n=3            \\  \hline
eigenvalue of $n\bar n$      &$626\sim636$ MeV &  $1317\sim1353$ MeV &$1872\sim1949$ MeV \\  \hline
eigenvalue of $s\bar s$      &   $830\sim848$ MeV &$1515\sim1544$ MeV      &    $2060\sim2130$ MeV     \\
\bottomrule[1pt]
\end{tabular*}
\\[1pt]
\footnotesize{$(a)$ }
  \end{minipage}
\\[1pt]
\begin{minipage}[!t]{\columnwidth}
  \renewcommand{\arraystretch}{1.3}
  \centering
  \setlength{\tabcolsep}{0.8mm}
\begin{tabular*}{130mm}{c@{\extracolsep{\fill}}cccccc}\toprule[1pt]
             &  $f_0(500)$             & $f_0(980)$      & $f_0(1370)$          & $f_0(1500)$            \\ \hline
mass         &$400\sim550$MeV          &$990\pm20$MeV    & $1200\sim1500$MeV    &$1506\pm6$MeV            \\ \hline
decay width  &$400\sim700$MeV         &$10\sim100$MeV    & $200\sim500$MeV       &$112\pm9$MeV          \\ \hline
             & $f_0(1710)$              &  $f_0(2020)$    & $f_0(2100)$                                  \\ \hline
mass         & $1704\pm12$ MeV   &$1992\pm16$ MeV    &$2101\pm7$MeV                                 \\ \hline
decay width  & $123\pm18$ MeV           &$442\pm60$MeV    &$224^{+23}_{-21}$MeV                           \\ \bottomrule[1pt]
\end{tabular*}
\\[1pt]
\footnotesize{$(b)$ }
 \end{minipage}
\caption{(a) The theoretically predicted mass spectra of $n\bar n$ and $s\bar s$ with the principal quantum numbers being n=1, 2 and 3, and (b) masses of the $f_0$ family which have been experimentally measured\cite{Tanabashi:2018oca}.
}
\label{tab1}
\end{table}

\par
As is supposed, $f_0(500)$ and $f_0(980)$ are mixtures of ground states of $n\bar n$ and $s\bar s$, thus we step forward to deal with the mixing of $n\bar n$ and $s\bar s$ to result in the physical eigenstates of $f_0(500)$ and $f_0(980)$.
The mixing matrix is written as
\begin{equation}
\begin{array}{rl}
\left(
  \begin{array}{ccc}
    m_{f_0(500)} & 0 \\
    0 & m_{f_0(980)}\\
      \end{array}
\right)
&=
\left(
  \begin{array}{ccc}
     \cos\theta & -\sin\theta \\
     \sin\theta & \cos\theta\\
      \end{array}
\right)
\left(
  \begin{array}{ccc}
     m_{\frac{u\bar u + d\bar d}{\sqrt 2}} & \lambda \\
     \lambda & m_{s\bar s} \\
      \end{array}
\right)
\left(
  \begin{array}{ccc}
     \cos\theta & -\sin\theta \\
     \sin\theta & \cos\theta\\
      \end{array}
\right)^{\dagger}\\
&=
\left(
  \begin{array}{ccc}
     \cos\theta & -\sin\theta \\
     \sin\theta & \cos\theta\\
      \end{array}
\right)
\left(
  \begin{array}{ccc}
     626\sim636MeV & \lambda \\
     \lambda & 830\sim848MeV\\
      \end{array}
\right)
\left(
  \begin{array}{ccc}
     \cos\theta & -\sin\theta \\
     \sin\theta & \cos\theta\\
      \end{array}
\right)^{\dagger}.
\end{array}
\end{equation}
Requiring $m_{f_0(500)}=400\sim550$ MeV and  $m_{f_0(980)}=990\pm20$ MeV, we find that when $\lambda=201\sim263$ MeV and $\theta=30.7^{\circ}\sim33.9^{\circ}$, our results coincide well with the conclusion of Refs.\cite{ElBennich:2008xy,Cheng:2002ai,Cheng:2005nb}.

\par
With our strategy, the five parameters of $\alpha_s(r)$, which is running with respect to $r$, $\kappa$, $c$ and $\lambda$, $\theta$ are determined, even though only certain ranges instead of exact numbers are provided.
It is believed that the results are in accordance with the experimental tolerance.

\par
It is noted that if one only considers the $q\bar q$ structure, the range from a few hundreds of MeV to 2 GeV can only accommodate six
P-wave $0^{++}$ eigenstates, so the masses of those excited eigenstates of $n\geq 3$ or $l\geq 3$ would be beyond this range.
There indeed exist seven $0^{++}$ physical mesons which are experimentally observed within the aforementioned range.
This fact signifies that there should exist something else beside the pure $q\bar q$ structures, and the most favorable candidate is mixtures of glueballs and $q\bar q$.
This observation inspires all researchers to explore the possible fractions of glueball components in the observed meson states.

\section{Study on the mixing of quarkonium and glueballs}
\par
Our work is  a phenomenological study and fully based on the available data.
As discussed in previous sections, we find that the energy region of $1\sim 2$ GeV cannot accommodate seven pure $0^{++}$ $q\bar q$ states, so the picture of pure $q\bar q$ structures is contrary to experimental observations.
Thus a scenario with a mixture of glueballs and $q\bar q$ within this energy region is favored.
The decay rates of $J/\psi\to \gamma+ f_0$ imply that $f_0(1370)$ and $f_0(1500)$ possess larger $q\bar q$ components, whereas $f_0(1710)$ has a large fraction of glueball component (see next section for detailed discussion).

\par
The lattice estimate suggests that the mass of the $0^{++}$ pure glueball is about 1.5 GeV, so one can naturally conjecture that
$f_0(1370)$, $f_0(1500)$ and $f(1710)$ are mixtures of $q\bar q$ and glueballs with certain fractions.
The rest of the $0^{++}$ $q\bar q$ states would have negligible probability to mix with glueballs because their masses are relatively far from that of the pure glueball.

\par
As an ansatz, we propose that the physical states $f_0(1370)$, $f_0(1500)$ and $f(1710)$ are mixtures of the second excited states of
$|N\rangle=n\bar n$ and $|S\rangle=|s\bar s\rangle$ with glueball state $|G\rangle$.
A unitary matrix $U$  transforms them into the physical states as
\begin{equation}
\begin{array}{rl}
\left(
  \begin{array}{ccc}
    |f_0(1370)\rangle \\
    |f_0(1500)\rangle \\
    |f_0(1710)\rangle \\
      \end{array}
\right) = U \left(
  \begin{array}{ccc}
    |N\rangle \\
    |S\rangle   \\
    |G\rangle    \\
  \end{array}
\right)
\end{array}
\label{pstoqs}
\end{equation}
and $U$ is a unitary matrix with the compact form
\begin{equation}
 U=\left(
  \begin{array}{ccc}
    c_{11} & c_{12} & c_{13} \\
    c_{21} & c_{22} & c_{23} \\
    c_{31} & c_{32} & c_{33}\\
      \end{array}
\right)
\end{equation}

\par
By imposing the unitary condition on $U$, we should determine all the elements of $U$ up to an arbitrary phase.
Furthermore we will enforce a few additional conditions on the shape of the matrix: (1) the determinant of the matrix must be unity, and (2) the matrix elements must be real.
Those requirements serve as a convention for fixing the unitary matrix.
The unitary matrix $U$  transforms the unphysical states $|N\rangle, |S\rangle$ and $|G\rangle$ into the physical eigenstates $|f_0(1370)\rangle, |f_0(1500)\rangle$ and $|f_0(1710)\rangle$, and at the same time diagonalizes the mass matrix $\tilde M$ as
\begin{equation}
\begin{array}{rl}
M_{f_0}=U \tilde MU^\dagger
\end{array}
\end{equation}
with
\begin{equation}
\begin{array}{rl}
M_{f_0}=
\left(
  \begin{array}{ccc}
    m_{f_0(1370)} & 0 & 0 \\
    0 & m_{f_0(1500)} & 0 \\
    0 & 0 & m_{f_0(1710)} \\
      \end{array}
\right)
\end{array}
\end{equation}
and
\begin{equation}
\begin{array}{rl}
\tilde M=
\left(
  \begin{array}{ccc}
    m_N & \lambda_{N-S} & \lambda_{N-G} \\
    \lambda_{N-S} & m_S & \lambda_{S-G} \\
    \lambda_{N-G} & \lambda_{S-G} & m_G \\
      \end{array}
\right)
\end{array}
\end{equation}
Namely, $m_{f_0(1370)},\: m_{f_0(1370)}$, and $m_{f_0(1710)}$ are the three roots of the equation
\begin{equation}
\begin{array}{rl}
&m_{f_0}^3 - m_{f_0}^2(m_G+m_S+m_N)+ m_{f_0}(m_G m_N + m_G m_S + m_N m_S - \lambda_{N-G}^2 - \lambda_{S-G}^2 - \lambda_{N-S}^2)\\
& - (\lambda_{N-G}^2 m_S + \lambda_{S-G}^2 m_N + \lambda_{N-S}^2 m_G
- 2\lambda_{N-G}\lambda_{S-G}\lambda_{N-S} - m_N m_S m_G)=0.
\end{array}
\end{equation}

\par
Since we know that QCD is flavor blinded, following Ref.\cite{Weingarten:1996pp}, the relation $\langle u\bar u/d\bar d|H|G\rangle=\langle s\bar s|H|G\rangle$ should be satisfied.
Thus we have
\begin{equation}
\begin{array}{rl}
\frac{\langle N|H|G\rangle}{\langle S|H|G\rangle}=
\frac{\langle \frac{u\bar u + d\bar d}{\sqrt 2}|H|G\rangle}{\langle s\bar s|H|G\rangle}=
\frac{ \frac{1}{\sqrt 2}+ \frac{1}{\sqrt 2} }{1}=\sqrt 2
\end{array}
\end{equation}
namely $\lambda_{N-G}=\sqrt 2 \lambda_{S-G}$.
Furthermore, the phase spaces and off-mass-shell quark effect may also affect the relation between $\lambda_{N-G}$ and $\lambda_{S-G}$, thus in our calculation we set the relation $\frac{\lambda_{N-G}}{\lambda_{S-G}}= 1.3\sim 1.5$.
Generally, we have three unknowns in the Hermitian matrix $\tilde M$: $\lambda_{N-S},\; \lambda_{S-G}$ and $m_G$.
There are three independent equations, so we can fix all of the three unknowns.
Moreover, the work of Close, Farrar and Li offers an opportunity to determine the relation between $c_{33}$ and $b_{1710}$ since we find that in
$f_0(1710)$ the glueball component is dominant (see next section), namely we have the relation
\begin{equation}
\begin{array}{rl}
b_{1710}\sim c_{33}^2 b(R[G]\to gg)\sim c_{33}^2\times 1.
\end{array}
\end{equation}
Carrying out the numerical computations, we obtain the transformation matrix which satisfies all the aforementioned requirements:
\begin{equation}
\begin{array}{rl}
U
=
\left(
  \begin{array}{ccc}
    -0.96\sim-0.87 & -0.21\sim-0.07 & -0.45\sim-0.25 \\
    0.14\sim0.41 & -0.94\sim-0.82 & -0.40\sim-0.30 \\
    -0.36\sim-0.17 & -0.53\sim-0.32 & 0.80\sim0.92 \\
      \end{array}
\right)
,
\end{array}
\label{transformation}
\end{equation}
With this transformation matrix, by solving the three mass equations, we obtain
\begin{equation}
\begin{array}{rl}
\tilde M
=
\left(
  \begin{array}{ccc}
     1276\sim1398.6MeV & -27\sim1MeV & -164\sim-89MeV \\
    -27\sim1MeV & 1526\sim1550MeV & -114\sim-63MeV \\
    -164\sim-89MeV & -114\sim-63MeV & 1570\sim1661MeV \\
      \end{array}
\right)
,
\end{array}
\label{mass}
\end{equation}

\par
The masses $m_{n\bar n}=1317\sim 1353$ MeV and $m_{s\bar s}=1515\sim 1544$ MeV in Tab.\ref{tab1}(a) are directly obtained by solving the relativistic Schr\"odinger equation in the section above.
However, for light quarkonia, the parameters $\alpha_s$ and $\kappa$ for the linear potential cannot be well determined, so we set a criterion
which involves a few physical inequalities to gain $m_{n\bar n}$ and $m_{s\bar s}$ within reasonable ranges.

\par
Then we input the two values of $m_{n\bar n}$ and $m_{s\bar s}$ into the non-diagonal mass matrix and by solving the secular equation we determine the masses of the physical $f_0$ states.
In principle we would simultaneously achieve the expected values of the non-diagonal matrix elements along with the physical masses of $f_0$.
However, we notice that the secular equation cannot be solved in the usual way, so we adopt an alternative method to simplify the problem.
We pre-determine the ranges of the elements of the unitary matrix which diagonalizes the mass matrix and then substitute them into the secular equation to check if the equation can be satisfied, if all the requirements (unitarity, etc.) are fulfilled.
Repeating the process many times, we find that the pre-determined ranges for the masses $m_{n\bar n}$ and $m_{s\bar s}$ of the first excited states of $n\bar n$ and $s\bar s$ should be shifted slightly, to $m_{n\bar n}=1276\sim 1398$ MeV and $m_{s\bar s}=1526\sim 1550$ MeV.
Obviously, the small shifts do not correspond to any quantitative changes, but indeed are identical, even though the superficial values look a bit different.
We can see that the newly obtained ranges roughly overlap with the previous ones.

\par
The solutions show that for $f_0(1370)$ and $f_0(1500)$, the main components are $q\bar q$ bound states, whereas the glueball component in $f_0(1710)$ is overwhelmingly dominant.
It also suggests the mass of a pure glueball of $0^{++}$ to be $1570\sim1661$ MeV.
This value is consistent with the results calculated in quenched lattice QCD: $1710\pm50\pm80$ MeV \cite{Chen:2005mg}, $1648\pm58$ MeV \cite{Vaccarino}, $1654\pm83$ MeV \cite{Loan} and $1622\pm29$ MeV \cite{Lee:1999}.

\section{ signal for glueball and light quark pair mixture in $f_0$ mesons }
\par
In this section we calculate the rates of  radiative decays $J/\psi\to \gamma+ f_0$ which may expose the structures of various $f_0$ states.

\subsection{ Determining fractions of glueball components in $f_0$ mesons via $J/\psi\to \gamma+ f_0$ decays }
\par
In this section let us briefly introduce the results of Close, Farrar and Li, without going into the details of the derivations.
In their pioneering work, it was proposed to determine the fraction of glueball components in a meson via  $J/\psi\to \gamma+ f_0$ decay.
We especially focus on mixtures of $f_0(1370)$, $f_0(1500)$ and $f_0(1710)$ states with glueballs because if they are pure quark-antiquark states, the theoretically estimated values of their mass spectra obviously deviate from the data (see above section).
In Ref.\cite{Close:1996yc,Cakir:1994jf}, for searching glueball fraction, an ideal reaction is the radiative decays of $J/\psi$.
Close, Farrar and Li formulated the decay branching ratios as
\begin{equation}
\begin{array}{rl}
R(J/\psi\to \gamma +f_0)=R(J/\psi\to \gamma +g g)\frac{c_R x|
H_{0^{++}}(x)|^2}{8\pi(\pi^2-9)}\frac{m_{f_0}}{m_\Psi^2}\Gamma(f_0\to g g),
\end{array}
 \label{branching}
\end{equation}
where $x=1-\frac{m_{f_0}^2}{m_\psi^2}$ and $c_R=2/3$ for the $0^{++}$ state, $H_{0^{++}}(x)$ is a loop integral and its numerical result
is given in Ref. \cite{Close:1996yc}.
The branching ratio $b$ is defined as
\begin{equation}
\begin{array}{rl}
b(f_0\to g g)=\frac{\Gamma(f_0\to g g)}{\Gamma(f_0\to all)}.
\end{array}
\label{ratio1}
\end{equation}
Taking experimental data\cite{Tanabashi:2018oca}, $BR(J/\psi \to \gamma gg)=(8.8\pm1.1)\%$, $BR(J/\psi \to \gamma f_0(1370) \to \gamma K\bar K)=(4.2\pm1.5) \times 10^{-4}$ and $BR(f_0(1370) \to K\bar K)=(35\pm13)\%$\cite{Bugg:1996ki}, and we can obtain the branching ratios of  $J/\psi \to \gamma f_0(1370)$ which are listed in Tab.\ref{tab2}.

\par
For $f_0(1500)$, combining $BR(f_0(1500)\to \pi\pi)=(34.5\pm2.2)\%$ and $BR(f_0(1500)\to \eta\eta)=(6.0\pm0.9)\%$ with $BR(J/\psi \to f_0(1500)\gamma \to \pi \pi \gamma)=(1.09\pm 0.24) \times 10^{-4}$ and $BR(J/\psi \to f_0(1500)\gamma \to \eta \eta \gamma)=(0.17\pm 0.14) \times 10^{-4}$\cite{Tanabashi:2018oca}, we have
\begin{equation}
\begin{array}{rl}
&BR(J/\psi \to \gamma f_0(1500))(\pi\pi)= 3.16\times10^{-4}(1\pm22.9\%) = (3.16\pm0.72)\times10^{-4}\\
&BR(J/\psi \to \gamma f_0(1500))(\eta\eta)= 2.83\times10^{-4}(1\pm83.7\%) = (2.83\pm2.37)\times10^{-4}
\end{array}
\end{equation}
namely
\begin{equation}
\begin{array}{rl}
&b_{1500}(\pi\pi)= 19.1\%(1\pm27.3\%)= (19.1\pm5.2)\%\\
&b_{1500}(\eta\eta)= 17.1\%(1\pm85.0\%)= (17.1\pm14.5)\%
\end{array}
\end{equation}
To consider possible experimental errors, in our later calculations, we use  $b_{1500}=0.171\pm0.145(\eta\eta)$ as the input.

\par
For $f_0(1710)$, there are two possible ways to get the branching ratio of $J/\psi\to f_0(1710)+\gamma$, and we adopt one of them, for which the experimentalists provide the following information on the four sequential channels\cite{Tanabashi:2018oca}
\begin{equation}
\begin{array}{rl}
&BR(J/\psi \to f_0(1710)\gamma \to K\bar K \gamma)=(9.5^{+1.0}_{-0.5})\times 10^{-4}\approx (9.5\pm 1.0) \times 10^{-4}\\
&BR(J/\psi \to f_0(1710)\gamma \to \eta \eta \gamma)=(2.4^{+1.2}_{-0.7})\times 10^{-4}\approx (2.4\pm 1.2) \times 10^{-4}\\
&BR(J/\psi \to f_0(1710)\gamma \to \pi \pi \gamma)=(3.8\pm 0.5) \times 10^{-4}\\
&BR(J/\psi \to f_0(1710)\gamma \to \omega \omega \gamma)=(3.1\pm 1.0) \times 10^{-4}\\
\end{array}
\end{equation}
as well the data on the decay modes of $f_0(1710)$\cite{Longacre:1986fh,Albaladejo:2008qa}:
\begin{equation}
\begin{array}{rl}
&BR(f_0(1710)\to K\bar K)=0.38^{+0.09}_{-0.19}\approx 0.38\pm 0.19;\\
&BR(f_0(1710)\to \pi \pi)=0.039^{+0.002}_{-0.024}\approx 0.039\pm 0.024;\\
&BR(f_0(1710)\to \eta \eta)=0.22\pm 0.12.\\
\end{array}
\end{equation}
Because these four channels probably dominate the radiative decay of $J/\psi\to f_0(1710)+\gamma$, as we summarize the four branching ratios, the resultant value should generally be close to unity.
A straightforward calculation determines $BR(J/\psi \to f_0(1710)\gamma)=(18.8\pm 1.9) \times 10^{-4}$ which corresponds to $b_{1710}=85.5\%(1\pm 21.8\%)=(85.5\pm18.6)\%$.

\par
An alternative approach is that we can directly use the available branching ratios of radiative decays of $J/\psi$ to do the same job.
\begin{equation}
\begin{array}{rl}
&BR(J/\Psi\to \gamma +f_0(1710))(K\bar K)=25.0\times 10^{-4}(1\pm 52.1\%)=(25.0\pm 13.0) \times 10^{-4}\\
&BR(J/\Psi\to \gamma +f_0(1710))(\pi\pi)=97.4\times 10^{-4}(1\pm 62.9\%) =(97.4\pm 61.3) \times 10^{-4}\\
&BR(J/\Psi\to \gamma +f_0(1710))(\eta\eta)=10.9\times 10^{-4}(1\pm 74.0\%) =(10.9\pm 8.1) \times 10^{-4}.
\end{array}
\end{equation}

\par
We have tried and noted that among the four channels, the calculated $BR(J/\Psi\to \gamma +f_0(1710))(\eta\eta)$ is too small.
The reason might originate from the error in measuring $\Gamma(f_0(1710)\to\eta\eta)$, whose value is not as reliable as the databook suggests\cite{pdglive}.
Thus we only use the first two results for calculating $b_{1710}$.
Then we obtain the corresponding $b_{1710}$ values as
\begin{equation}
\begin{array}{rl}
&b_{1710} (K\bar K)=1.14(1\pm 54.6\%)=1.14\pm 0.62\\
&b_{1710} (\pi\pi)=4.43(1\pm 65.8\%)=4.43\pm 2.91.
\end{array}
\end{equation}
For $b_{1710} (K\bar K)$, even though the superficial central value of the $b$ factor is above 1.0, when a large experimental error is taken into account, it is comparable with the value of $b_{1710}=(85.5\pm18.6)\%$.
The two values are reasonably consistent with each other and this almost confirms that the aforementioned sequential decay channels dominate the radiative decays of $J/\psi\to f_0(1710)+\gamma$.
Thus for $BR(J/\psi \to f_0(1710)\gamma)$ we take its experimental value as $(18.8\pm 1.9) \times 10^{-4}$.

\par
We then obtain all the results which are listed in Tab.\ref{tab2}.
\begin{table}[htbp]
\begin{center}
\footnotesize
\begin{tabular*}{90mm}{c@{\extracolsep{\fill}}cccc}\hline
            &  $BR(J/\psi\to \gamma f_0)$     & $b(f_0\to g g)$    \\\hline
$f_0(1370)$   & $(12.0\pm6.2)\times10^{-4}$\cite{Tanabashi:2018oca,Bugg:1996ki}  & $27.5\pm19.4\%$           \\
$f_0(1500)$   & $(2.8\pm2.4)\times10^{-4}$\cite{Tanabashi:2018oca}  & $17.1\pm14.5\%$           \\
$f_0(1710)$   & $(18.8\pm1.9)\times10^{-4}$\cite{Tanabashi:2018oca}  & $85.5\pm18.6\%$        \\\hline
\end{tabular*}
\caption{  $b(f_0\to g g)$ for $f_0(1370)$, $f_0(1500)$ and $f_0(1710)$ from experimental data. }
\label{tab2}
\vspace{0mm}
\end{center}
\end{table}

\par
\begin{figure}[htbp]
\begin{center}
\includegraphics[width=0.5\textwidth]{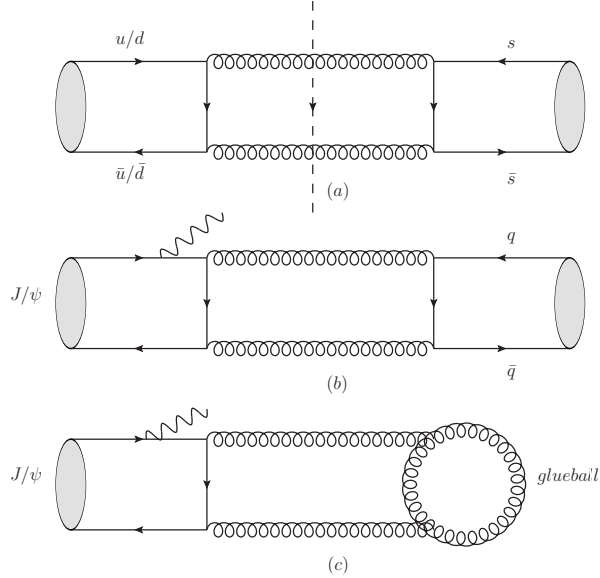}
\caption{
(a) Mixing between $n\bar n$ and $s\bar s$, (b) $J/\psi\to \gamma+ f_0$ with $f_0$ as a $q \bar q$ bound state, and (c) $J/\psi\to \gamma+ f_0$ with $f_0$ as a glueball.
}
\label{guleball}
\end{center}
\end{figure}

As indicated in Ref.\cite{Close:1996yc}, the width of $f_0$ is determined by the inclusive processes of $f_0\to gg$ and $f_0\to q\bar q$.
It is noted that the contribution of the glueball component to the $gg$ final state is of order 1, as $b(R[G]\to gg)\sim 1$, whereas
\begin{equation}
\begin{array}{rl}
b(R[q\bar q]\to gg)=\mathcal{O}(\alpha^2_s)\approx 0.1\sim0.2.
\end{array}
\label{ratio}
\end{equation}
In Fig.\ref{guleball} we show that the value of $b(R[q\bar q]\to gg)$ is different if $f_0$ is a glueball or $q\bar q$ bound state; readers can ignore the irrelevant hadronization processes.

\par
Combining with the updated experimental data, one can conclude from the pioneer paper of Close et al. that $f_0(1370)$ and $f_0(1500)$ possess larger $q\bar q$ components whereas $f_0(1710)$ has a large fraction of glueball consituent.

\subsection{rates of $f_0$ decaying into two pseudoscalar mesons}
\par
Following Refs.\cite{Cheng:2006hu,Cheng:2015iaa}, we have
\begin{equation}
\begin{array}{rl}
&|\mathcal{M}(F_i\to K\bar K)|^2= 2f_1^2(r_a \frac{c_{i1}}{\sqrt 2}+ c_{i2}+ 2g_s^{K\bar K}c_{i3})^2\\
&|\mathcal{M}(F_i\to \pi\pi)|^2 = 6f_1^2(\frac{c_{i1}}{\sqrt 2}+ g_s^{\pi\pi}c_{i3})^2\\
&|\mathcal{M}(F_i\to \eta\eta)|^2 = 2f_1^2(a_\eta^2 \frac{c_{i1}}{\sqrt 2}+ r_a b_\eta^2 c_{i2}+ g_s^{\eta\eta} (a_\eta^2+ b_\eta^2) c_{i3}+
g_{ss} (2a_\eta^2+ b_\eta^2+ \frac{4}{\sqrt2}a_\eta b_\eta)c_{i3})^2
\end{array}
\end{equation}
where$F_i=(f_0(1370),f_0(1500),f_0(1710))$, $f_1$ is the coupling constant of the OZI-allowed Feynman diagrams defined in Ref.\cite{Cheng:2006hu}, $g_s$ is the ratio of the OZI-suppressed coupling constant to that of the OZI-allowed one, $g_{ss}$ is the ratio of the doubly OZI-suppressed coupling constant to that of the OZI-allowed one, and $r_a$ denotes a possible $SU(3)$ breaking effect in the OZI allowed decays\cite{Cheng:2006hu}.
For $a_\eta$ and $b_\eta$ we have the relations
\begin{equation}
\begin{array}{rl}
&a_\eta=\frac{\cos\theta- \sqrt2\sin\theta}{\sqrt3}\\
&b_\eta=-\frac{\sin\theta+ \sqrt2\cos\theta}{\sqrt3}
\end{array}
\end{equation}
with $\theta=-14.4^{\circ}$\cite{Cheng:2006hu}.
References\cite{Sexton:1995kd,Burakovsky:1998zg} show the relations $g_s^{\pi\pi}:g_s^{K\bar K}:g_s^{\eta\eta}=0.834^{+0.603}_{-0.579}:2.654^{+0.372}_{-0.402}:3.099^{+0.364}_{-0.423}$ through lattice calculation.
In Ref.\cite{Cheng:2006hu} the authors take two schemes with $g_s^{K\bar K}/g_s^{\pi\pi}=1.55$ for scheme I and $3.15$ for scheme II.
Then for scheme I they take $g_s^{\pi\pi}:g_s^{K\bar K}:g_s^{\eta\eta}=1:1.55:1.59$ and after fitting data obtain $g_s^{\pi\pi}=-0.48$, $g_{ss}=0$ and $r_a=1.21$.
For scheme II they took $g_s^{\pi\pi}:g_s^{K\bar K}:g_s^{\eta\eta}=1:3.15:4.74$ and obtained $g_s^{\pi\pi}=0.10$, $g_{ss}=0.12$ and $r_a=1.22$. During their fitting process, $m_{s\bar s}$ and $m_G$ as input parameters were set in the same ranges as in our article, but for the value of $m_{n\bar n}$ they took an input parameter $100$MeV larger than that in our article.
Considering the uncertainty of $g_s^{\pi\pi}:g_s^{K\bar K}:g_s^{\eta\eta}$, the two sets of parameters do not have a qualitative difference.

\par
From Refs.\cite{Cheng:2006hu,Cheng:2015iaa} we have
\begin{equation}
\begin{array}{rl}
R^{F_i}_{\pi\pi/K\bar K}=\frac{\Gamma(F_i \to \pi\pi)}{\Gamma(F_i \to K\bar K)}=
3\frac{(\frac{c_{i1}}{\sqrt 2}+ g_s^{\pi\pi}c_{i3})^2}{(r_a \frac{c_{i1}}{\sqrt 2}+ c_{i2}+ 2g_s^{K\bar K}c_{i3})^2}\frac{p_\pi}{p_K}.
\end{array}
\end{equation}
Then we list the ratios of $F_i\to PP$ for our predictions and experimental data in Tab.\ref{tab3}.
\begin{table}[htbp]
\begin{center}
\renewcommand\arraystretch{1.5}
\begin{tabular}{cccc}
\hline
\hline
 ~~ & ~~experimental value~~  &  ~~scheme I~~ & ~~scheme II~~  \\
\hline
$R^{f_0(1370)}_{\pi\pi/K\bar K}$       &  $>1$   &  $3.98\sim11.68$ &  $1.22\sim1.88$   \\
$R^{f_0(1370)}_{\eta\eta/\pi\pi}$        &    &  $0.01\sim0.05$ &  $0.12\sim0.21$   \\
$R^{f_0(1370)}_{K\bar K/\eta\eta}$      &    &  $5.53\sim7.44$ &  $3.84\sim4.31$   \\
$R^{f_0(1500)}_{\pi\pi/K\bar K}$      &  $4.1\pm0.5$ &  $1.81\sim9520.07$ &  $0.02\sim0.47$   \\
$R^{f_0(1500)}_{\eta\eta/\pi\pi}$   &  $0.173\pm0.024$ &  $8.65\times10^{-5}\sim0.14$ &  $0.78\sim17.32$   \\
$R^{f_0(1500)}_{K\bar K/\eta\eta}$      &  $1.43\pm0.24$   &  $0.48\sim10.72$ &  $2.71\sim3.31$   \\
$R^{f_0(1710)}_{\pi\pi/K\bar K}$  &  $0.23\pm0.05$ &  $0.30\sim0.41$ &  $0.71\sim112.38$   \\
$R^{f_0(1710)}_{\eta\eta/\pi\pi}$   &  $2.09\pm0.80$ &  $0.59\sim0.82$ &  $9.69\times10^{-4}\sim4.87$   \\
$R^{f_0(1710)}_{\eta\eta/K\bar K}$  &  $0.48\pm0.15$ &  $0.24\sim0.25$ &  $6.86\times10^{-4}\sim114.20$   \\
\hline
\hline
\end{tabular}
\caption{
ratios of $F_i\to PP$ for our predictions in scheme I and II, and for experimental data. All the experimental data for $R^{f_0(1370)}_{\pi\pi/K\bar K}$ are taken from the PDG\cite{Tanabashi:2018oca}. In the PDG the range of $R^{f_0(1370)}_{K\bar K/\pi\pi}$ varies from $0.08\pm0.08$ to $0.91\pm0.20$, thus we only consider $R^{f_0(1370)}_{\pi\pi/K\bar K}>1$ here.
}
\label{tab3}
\end{center}
\end{table}

\par
From the table above we can see that in scheme I the upper limit of $R^{f_0(1500)}_{\pi\pi/K\bar K}$ is $9520.07$.
This is caused by the destructive interference between the contributions of the glueball and quarkonia ingredients to $f_0(1500)\to K\bar K$; the lower bound of $R^{f_0(1500)}_{\eta\eta/\pi\pi}$ is tiny because of a destructive interference between the glueball and quarkonia contributions to $f_0(1500)\to \eta\eta$.
For scheme II, the destructive interference between the glueball and quarkonia contributions to $f_0(1710)\to \eta\eta$ and $K\bar K$ lead the lower bound of $R^{f_0(1710)}_{\eta\eta/\pi\pi}$ and $R^{f_0(1710)}_{\eta\eta/K\bar K}$ to be negligible and the upper limit of $R^{f_0(1710)}_{\pi\pi/K\bar K}$ and $R^{f_0(1710)}_{\eta\eta/K\bar K}$ to be as large as $10^2$.
Different from Refs.\cite{Cheng:2006hu,Cheng:2015iaa} where the authors preferred scheme II, we find that in our calculation scheme I may fit the experimental data better.
With scheme I, our theoretical prediction for the ratios of the decay rates for the three channels of $f_0(1500)$ ($R^{f_0(1500)}_{\pi\pi/K\bar K}$, $R^{f_0(1500)}_{\eta\eta/\pi\pi}$ and $R^{f_0(1500)}_{K\bar K/\eta\eta}$) deviate only slightly from experimental data, while for scheme II, the theoretical predictions obviously deviate from experimental data.
Cheng et al. considered the two schemes based on the lattice results $g_s^{\pi\pi}:g_s^{K\bar K}:g_s^{\eta\eta}=
0.834^{+0.603}_{-0.579}:2.654^{+0.372}_{-0.402}:3.099^{+0.364}_{-0.423}$.
Considering the large uncertainty, we believe our theoretical predictions can fulfill the experimental constraints.

\subsection{rates of $f_0$ decaying into two photons}

\par
The $f_0(1710)\to \gamma\gamma$ decay would be the most sensitive channel to test the glueball fraction inside the hadron because gluons
do not directly couple to photons.
In early searches for glueballs, the rate of  possible glueball decay into photons was considered to be seriously depressed and the mechanism  was described by the word ``stickiness," which was the first criterion to identify a glueball.
Thus for $f_0\to \gamma\gamma$, the amplitude is
$$M(f_0\to\gamma\gamma)=c_n<\gamma\gamma|H_{eff}|N>+c_s <\gamma\gamma|H_{eff}|S>+c_G<\gamma\gamma|H'_{eff}|G>,$$
and because $H_{eff}'$ is a loop-induced effective Hamiltonian, it suffers an $\mathcal{O}(\frac{\alpha_s}{\pi})$ suppression\cite{Ahrens:2011px} compared to $H_{eff}$.
A detailed computation of the box-diagram has been given in the literature, but here we just make an order of magnitude estimate, which is enough for the present experimental accuracy.

\par
With this principle, here let us make a prediction of $F_i\to \gamma\gamma$ for $f_0(1370)$, $f_0(1500)$ and $f_0(1710)$ based on the values we have obtained in this work.
As a comparison we list what the authors of Ref.\cite{Cheng:2015iaa} predicted
\begin{equation}
\begin{array}{rl}
\Gamma(f_0(1370) \to \gamma\gamma):\Gamma(f_0(1500) \to \gamma\gamma):\Gamma(f_0(1710) \to \gamma\gamma)= 9.3:1.0:1.7
\end{array}
\end{equation}
through the relation
\begin{equation}
\begin{array}{rl}
\Gamma(F_i \to \gamma\gamma)\sim (\frac{5}{9}\frac{c_{i1}}{\sqrt2}+ \frac{1}{9}c_{i2})^2
\end{array}
\end{equation}
where they neglected the contribution of the glueball component in $F_i$ due to the $\mathcal{O}(\frac{\alpha_s}{\pi})$ suppressed contribution to the amplitude.
Although the contribution from the glueball component is suppressed, it may reach the error range determined by accurate measurement, thus in our calculation we present the ratios of $F_i\to \gamma\gamma$ with and without considering the contribution of the glueball:
\begin{equation}
\begin{array}{rl}
\Gamma(F_i \to \gamma\gamma)\sim |\frac{1}{\sqrt3}(\frac{5}{9}\frac{c_{i1}}{\sqrt2}+ \frac{1}{9}c_{i2})+
\frac{Tr[t^a t^a]}{\sqrt8}\mathcal{O}(\frac{\alpha_s}{\pi})\frac{6}{9}c_{i3}|^2.
\end{array}
\end{equation}
In our numerical calculations we take $\alpha_s\sim 0.3$ as a example.
We list the correspond results in Tab.\ref{tab4}.
\begin{table}[h]
\begin{center}
\renewcommand\arraystretch{1.5}
\begin{tabular}{cccc}
\hline
\hline
 ~~ & ~~results from Ref.\cite{Cheng:2015iaa}~~&~~without glueball component~~ &  ~~with glueball component~~ \\
\hline
$\frac{\Gamma(f_0(1370) \to \gamma\gamma)}{\Gamma(f_0(1500) \to \gamma\gamma)}$ & 9.3 & $27.2\sim1.93\times10^7$ & $19.0\sim3208.3$ \\
$\frac{\Gamma(f_0(1710) \to \gamma\gamma)}{\Gamma(f_0(1500) \to \gamma\gamma)}$ & 1.7 &$5.8\sim4.13\times10^6$ &  $0.003\sim33.0$   \\
$\frac{\Gamma(f_0(1710) \to \gamma\gamma)}{\Gamma(f_0(1370) \to \gamma\gamma)}$ & 0.183 & $0.087\sim0.261$ &  $6.8\times10^{-5}\sim0.016$   \\
\hline
\hline
\end{tabular}
\caption{
Ratios of $F_i\to \gamma\gamma$ with and without considering the contribution from the glueball component.
}
\label{tab4}
\end{center}
\end{table}

\par
From the table above we find that without considering the contribution from the glueball component, the upper limits of $\Gamma(f_0(1370) \to \gamma\gamma)/\Gamma(f_0(1500) \to \gamma\gamma)$ and $\Gamma(f_0(1710) \to \gamma\gamma)/\Gamma(f_0(1500) \to \gamma\gamma)$ are extremely large, because of a destructive interference between the $n\bar n$ and $s\bar s$ contributions to $f_0(1500)\to \gamma\gamma$.
Taking into account the contribution from the glueball component, the lower limit of $\Gamma(f_0(1710) \to \gamma\gamma)/\Gamma(f_0(1370) \to\gamma\gamma)$ is tiny.
This is caused by the destructive interference between the contributions of the glueball and quarkonia ingredients to $f_0(1710)\to \gamma\gamma$, whereas the upper bound of $\Gamma(f_0(1370)\to \gamma\gamma)/\Gamma(f_0(1500)\to\gamma\gamma)$ is very large, which is caused by the destructive interference between $n\bar n$ and $s\bar s$ contributions to $f_0(1500)\to\gamma\gamma$.

\par
Apart from these extreme cases, we find that the decay width of $f_0(1710) \to \gamma\gamma$ is smaller than that of $f_0(1370) \to \gamma\gamma$ by one or two orders of magnitude for our structure assignments, i.e in $f_0(1710)$ the glueball component is dominant while in $f_0(1370)$ the quark component is dominant.
The decay width of $f_0(1370) \to \gamma\gamma$ is larger than that of $f_0(1500) \to \gamma\gamma$ due to the fact that the $n\bar n$ and $s\bar s$ components constructively interfere for $f_0(1370)$ whereas they destructively interfere for $f_0(1500)$ in our scenario.
The prediction is somewhat different from that made by Cheng et al\cite{Cheng:2015iaa}.
The comparison is shown in the above tables.

\section{discussion and conclusion}
\par
The main purpose of this work is to explore the probability of mixing between $0^{++}$ $q\bar q$ states and glueballs.
To serve this goal, we first calculated the mass spectra of six $0^{++}$ light $q\bar q$ bound states by solving the relativistic Schr$\ddot{o}$dinger equation.

\par
The numerical estimates indicate that in order to fit the observed experimentally measured spectra of $f_0(500)$, $f_0(980)$, $f_0(1370)$, $f_0(1500)$, $f_0(1710)$, $f_0(2020)$ and $f_0(2100)$, an extra hadronic structure is needed to accommodate the seven members of the $f_0$ family
existing in the energy range from a few hundreds of MeV to 2 GeV.
As suggested in the literature, the most favorable scenario is the mixing between $q\bar q$ and glueballs of the same quantum numbers.
Instead of calculating the mixing based on complete theoretical frameworks, we investigate the mixing by analyzing experimental data.
Besides properly diagonalizing the mass matrix, supplementary information about the fractions of the glueball components in the $f_0$ mesons can be extracted from the data of $J/\psi$ radiative decays to $f_0$.
It is found that in $f_0(1370)$ and $f_0(1500)$ there are mainly $q \bar q$ bound states whereas in $f_0(1710)$ a glueball component dominates.

\par
In this work, we obtained the mixing parameters by a phenomenological study, while some authors have tried to calculate them directly in terms of certain models.
Within this energy range, the dominant dynamics is the non-perturbative QCD which induces the mixing.
Since solid knowledge about non-perturbative QCD is still lacking, the theoretical calculation heavily relies on the models adopted, where some model-dependent parameters have to be input and cause uncertainties in the theoretical estimates.
Among those calculations, the results of the lattice calculations \cite{Morningstar:1999rf, Chen:2005mg, Bali:1993fb} and those based on the QCD sum rules \cite{Narison:1996fm, Huang:1998wj, Shuiguo:2010ak, Harnett:2000fy, Forkel:2003mk} may make more sense even though still not completely trustworthy.
Combining the phenomenological studies by analyzing the experimental data and those estimates based on theoretical frameworks may shed light on this intriguing field.

\par
Now let us briefly discuss the other $0^{++}$ states $f_0(500)$, $f_0(980)$, $f_0(2020)$ and $f_0(2100)$.
Since their masses are far below or above the assumed glueball mass, according to the principles of quantum mechanics, their mixing with glueballs should be to be small and can be ignored at the first order of approximation.
In our calculation, $f_0(500)$ and $f_0(980)$ are considered as mixtures of ground states of $n\bar n$ and $s\bar s$.
This result is consistent with the conclusion of Refs.\cite{Cheng:2002ai,Cheng:2005nb}.
Alternatively, in Ref.\cite{Fariborz} the authors studied the five $0^{++}$ states $f_0(500)$, $f_0(980)$, $f_0(1370)$, $f_0(1500)$ and $f_0(1710)$, concluding that those five states are composed of two lowest-lying four-quark scalar meson nonets, two next-to-lowest lying two-quark nonets, and a scalar glueball.
In their work, $f_0(500)$ is considered as a non-strange four-quark component dominated bound state rather than a quark pair bound state.

\par
The reason that we are able to carry out this exploration is that much more data in this energy range has been accumulated and the measurements are obviously more accurate than 25 years ago.
However, as one can see, the precision is still far below the requirement for determining the mixing parameters well.
We therefore set our hope on the experimental progress which will be made by the BESIII, BELLE and LHCb experiments, and probably the future charm-tau factory.
To verify this mixing scenario one certainly needs to do more theoretical work, including estimating the production (not only via the radiative decays of $J/\psi$) and decay rates of $f_0$ families.
Further work, both experimental and theoretical is badly needed.

\par
Moreover, in this work, following the strategy provided by Close et al., we suppose that the $f_0$ family only contains mixtures of light quarkonia and glueballs and have carried out calculations on the mass spectra of the mixtures.
Obviously the phenomenological consequences depend heavily on the ansatz.
As we state above, it is probably true for the first order approximation.
It has been noted that the $f_0$ family may not only be mixtures of glueballs and light quakonia, but also hybrids made of $q\bar qg$\cite{He:2006tw,He:2006ij} or even four-quark states\cite{Li:2013nba,Li:2006ru}.
Therefore, we are not going to make a bold prediction here, but as promised, we will redo the estimates which were done in Refs.\cite{He:2006tw,He:2006ij} and \cite{Li:2013nba,Li:2006ru}, based on the new framework.
We may then provide some theoretical predictions for the decay rates of $f_0$ mesons which can soon be checked by more accurate data from BESIII, BELLE and LHCb soon.

\vskip 10mm
\par
\noindent{\large\bf ACKNOWLEDGMENTS} \\
This work is supported by National Natural Science Foundation of China under the Grant Number 11805160, 11747040, 11675082, 11605039.

\end{document}